\documentclass[preprintnumbers, prd, onecolumn, showpacs,floatfix,preprintnumbers,
superscriptaddress, nofootinbib]{revtex4}
\usepackage{graphicx}
\usepackage{epsfig}
\usepackage{bm}
\usepackage{amssymb}
\usepackage{float}
\usepackage{amsmath}
\usepackage{dcolumn}
\usepackage{cancel}
\usepackage[colorlinks]{hyperref}
\usepackage[usenames,dvipsnames]{color}
\hypersetup{
     breaklinks=true,
    pdfstartview={FitH},    
    colorlinks=true,       
    linkcolor=blue,          
    citecolor=red,        
    filecolor=magenta,      
    urlcolor=blue,           
    anchorcolor=green,      
    linktocpage=true
}

\def\doi{http://doi.org}

\begin{document}

\title{Black Shells, Dirac's Field and the species problem}

\author{W. A. Rojas C}
\email{warojasc@unal.edu.co}
\affiliation{Departamento de F\'isica, Universidad Nacional de Colombia,  Bogot\'a, Colombia}

\author{J. R. Arenas S.}
\email{jrarenass@unal.edu.co}
\affiliation{Observatorio Astron\'omico Nacional\\ Universidad Nacional de Colombia, Bogot\'a, Colombia}

\newcommand{\be}{\begin{equation}}
\newcommand{\ee}{\end{equation}}
\newcommand{\bea}{\begin{eqnarray}}
\newcommand{\eea}{\end{eqnarray}}
\newcommand{\bc}{\begin{center}}
\newcommand{\ec}{\end{center}}
\begin{abstract}
We  describe  a thermal atmosphere around  a black hole as vacuum excitations near to gravitational radius of a  contracting thin black shell, i.e., in terms of properties of the physical vacuum  of fields around a thin shell of  mass $M$ collapsing from  infinity to the Schwarzschild radius according to an external stationary observer.

A natural explanation is introduced for the necessary cutoff using the equations of motion of the shells.

We make a thermodynamic description of a fermionic field near the gravitational radius. Then a  solution to the species  problem for  two fields, scalar one and spinor one, is proposed.

Finally we get the Bekenstein-Hawking entropy  as entanglement entropy of a thermal atmosphere, independent from number of fields.
\end{abstract} 
\maketitle 
\section{Introduction}
Thermodynamics describes the current behavior of physical systems with a high amount of particles. This conduct is well-structured along the statistical physics basis.
On the other hand, it is well known that a black hole has got analogical properties towards the standard thermodynamics. This black holes thermodynamic relies on some physical guidelines according to  the no-hair theorem and the cosmic sensorship hypotesis. So that, the black holes thermodynamics  laws and the generalized second law of thermodynamics,  compose a general framework where the entropy has a very important meaning \cite{BJ, HS}

	\[S_{BH}=\frac{ k_{B}c^{3}A}{4\pi\hbar G},
\]
where  $S_{BH}$  belongs to the Bekenstein-Hawking entropy and $A=A_{H}$ is the horizon area.

Bekenstein and Hawking showed that the entropy was equal to a quarter of a black hole event horizon area.

Hawking showed that black holes have thermal radiation with temperature produced by
	\[T_{H}=\frac{\hbar \kappa_{0}}{2\pi k_{B}c},
\]

where $\kappa _{0}$, is the surface gravity on the black hole horizon. In accordance with the aforementioned, the thermal radiation emitted by a black hole is called  Hawking radiation which has got a specific temperature known as Hawking temperature  \cite{HS}.

Within the scientific community, there is the strong conviction that the $S_{BH}$ has explanation in  context of quantum gravity. So like the  standard thermodynamics has explanation in statistical physics. If Boltzmann is  right
	\[S=\ln\left|\Omega\right|,
\]
where, the entropy is equal to the quantity of microstates of system $\Omega$. This group of microstates is related to the degrees of freedom accessible in a black hole. Such microscopic description seems to require a quantum gravity theory. However, in this paper we show for an external observer that it is not necessary to know details of quantum gravity to calculate $S_{BH}$.

An additional question arises in black hole studies: can this kind of systems have a loss of  information? According to Hawking's statements: 
 If a  black hole is product of gravitational  collapse, the quantum fields evolution, do not keep the unitarity due the black hole evaporation process. This fact suggests  that a loss of information can be a place in the black hole evaporation process.

In this context, there are  several explanations for $S_{BH}$ of black  holes. For instance, Gibbons and  Hawking approximation, where the  entropy is  associated with the spacetime topology \cite{GG}.  In the  brick wall model of 'thooft, the  entropy is associated to statistical entropy of  a warm quantum field close to the horizon \cite{tG}. Another candidate for this explanation  of $S_{BH}$, is the entanglement entropy. Entanglement  entropy is related to the modes and the hide correlations of  external observer when they are close to the event horizon \cite{AJ}.

Along the research of the entanglement entropy,  Fursaev identifies the following matters \cite{FD}
\begin{enumerate}
	\item The  entropy $S$ depends on the  cutoff $\epsilon$. Therefore, there must be some natural explanation why it should be adjusted for $S=S_{BH}$.
	\item In general $S$, receives contributions from all quantum fields in the nature. It depends on  the whole amount of field an their spins. The effective number $N$ of basic quantum fields is on the order $300$ \cite{MS}. However, $S_{BH}$ does not such dependence.
\end{enumerate}

The interpretation of entanglement state is related with the brick wall model of 'thooft. We take this model as black shell, that is, a thin spherical shell which contracts from the infinity to the Schwarzschild radius.  This shows that thermal atmosphere arises from the excitations of the  fundamental state of Boulware. Moreover an exactly localization of these excitations is given the Hartle-Hawking state. \cite{FD}.

This study is organized as following: first, we review the cinematic of Shell, the motion equation and we compares the  solution proposed by Israel \cite{IW2}, Arenas- Castro \cite{JR} and Akhmedov-Godazgar-Popov \cite{AE}. Also their consequences are analyzed.

A second part, is dedicated to find  the proper altitude  above the horizon, $\alpha$ and we calculate the thermodynamics properties for a fermionic field ($s=1/2$) close to the Shell.

Finally, the last part is about the species problem for entanglement entropy. The total  entropy for scalar and  fermionic fields is  equal to 
	\[S_{\mbox{F}}+S_{\mbox{B}}=S_{BH}.
\]
This result suggest, that the $S_{B}$ is independent to number of quantum fields, their spins and internal interactions \cite{JT,CY}. 
\section{What is   a Black Shell?}
The sight of an external observer in a far asymtotic region, includes a thin spherical shell of dust of mass $M$ which has gravitational contraction from infinity  to the Schwarzschild radius $r_{s}=2M$. In  the external region of the thin Shell, the spacetime is Schwarzschild like \cite{PF}

So that, we have that the   shell's surface  is a hypersurface spacelike $\Sigma \subset \mathcal{M} $which superficial stress-energy tensor is        \cite{PF, PE, IW, CS, IW2, JF, KJ, CRR,BI,GO}
\begin{equation}
S^{ab}=\sigma u^{a}u^{b},
\end{equation}

where $\sigma$ is the matter-energy density over the shell.
\subsection{Kinematics of hypersurfaces }
We have  a $(3+1)$ manifold where there is a hypersurface $\Sigma$, with the  condition $\Sigma\subset \mathcal{M}$ and $\Sigma$ can be either timelike, spacelike, or null. A specific hypersurface $\Sigma$ can be selected, when we restrict the  coordinates  $x^{\alpha}\in \mathcal{M} $ as
\begin{equation}
\Phi\left(x^{\alpha}\right)=0
\end{equation}
and parametric equations as
\begin{equation}
x^{\alpha}=x^{\alpha}(y^{a}),\,\,\,x^{\alpha}\in \mathcal{M},\,\,\,y^{a}\in \Sigma\,\,\,\mbox{and}\,\,\,\Sigma\subset \mathcal{M},
\label{A1}
\end{equation}
where $y^{a}$ is intrinsic coordinates of $\Sigma$.

Additional, $\Sigma$ is characterized for a normal vector $n _{\alpha}$. This vector is unitary and is 
\begin{align}
n^{\alpha}n_{\alpha}&=\epsilon
	\notag\\
&    \hspace{0.0cm}
=-1,\,\,\,\mbox{If $\Sigma$ is spacelike.}
	\notag\\
&    \hspace{0.0cm}
=1,\,\,\,\mbox{If $\Sigma$ is timelike.}
\label{A3}
\end{align}
where
\begin{equation}
n_{\alpha}=\frac{\epsilon \partial_{\alpha }\Phi}{\sqrt{g^{\mu\nu}\partial_{\mu }\Phi\partial_{\nu }\Phi}}.
\end{equation}
	
\subsection{First fundamental form: induced metric}
The induced metric is obtained when movements are limited on the hypersurface $\Sigma$. Such metric is 
\begin{equation}
ds^{2}_{\Sigma}=h_{ab}dy^{a}dy^{b}, \,\,\,h_{ab}=g_{\alpha \beta}e^{\alpha}_{a}e^{\beta}_{b},
\label{A4}
\end{equation}
where $h_{ab}$ is known as the  induced metric or first fundamental form and tangent vectors to the integral curves in $\Sigma$ are
\begin{equation}
e^{\alpha}_{a}=\frac{\partial x^{\alpha}}{\partial y^{a}}.
\label{A5}
\end{equation}
\subsection{Second fundamental form: extrinsic curvature}
The intrinsic curvature of manifold  $\mathcal{M}$  is determined for Riemann tensor, $R^{\alpha}_{\beta\gamma\delta}$
\begin{equation}
R^{\alpha}_{\beta\gamma\delta}=\frac{\partial \Gamma^{\alpha}_{\delta\beta}}{\partial x^{\gamma}}+\frac{\partial \Gamma^{\alpha}_{\gamma\beta}}{\partial x^{\delta}}+\Gamma^{\alpha}_{\gamma\sigma}\Gamma^{\sigma}_{\delta\beta}-\Gamma^{\alpha}_{\sigma\delta}\Gamma^{\sigma}_{\delta\beta}.
\end{equation}
The  extrinsic curvature or second fundamental form $K_{ab}$, defines how is curved $\Sigma$ hypersurface related to  $\mathcal{M}$, it is embedded.
So, $K_{ab}$ is related with normal derivative of metric  tensor $g_{\alpha\beta}$
\begin{equation}
K_{ab}=\frac{1}{2}\left[\mathcal{L}_{n}g_{\mu\nu}\right]e^{\alpha}_{a}e^{\beta}_{b}=n_{\alpha;\beta} e^{\alpha}_{a}e^{\beta}_{b}, 
\label{AW6}
\end{equation}

\subsection{ Formalism of Darmois-Israel }
Being the hypersurface $\Sigma$, it divides   spacetime in two regions $M^{+}$ and $M^{-}$, such that $g^{+}_{\alpha\beta}\in M^{+}$ and $g^{-}_{\alpha\beta}\in M^{-}$

\begin{figure}[h]
\centering
	\includegraphics[width=0.7 \textwidth]{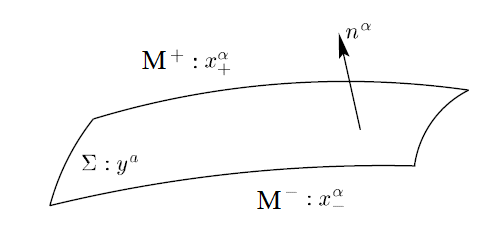}
\caption{Two regions of spacetime that joined in a common boundary. The  Figure is take from \cite{PE}.}
\label{}
\end{figure}
\subsection{The first condition of junction}
The first condition of junction states: the induced metric in both sides of  
$\Sigma$ is the same one. We can take it, as the continuity of first fundamental form
\begin{equation}
\left[h_{ab}\right]=h_{ab}^{+}|_{\Sigma}-h_{ab}^{-}|_{\Sigma}=0.
\label{A7}
\end{equation}

\subsection{Second condition of junction}
This condition declares that the extrinsic curvature is the same in both sides of hypersurface $\Sigma$. This can be interpreted as the continuity of the second fundamental  form
\begin{equation}
\left[K_{ab}\right]=K_{ab}^{+}|_{\Sigma}-K_{ab}^{-}|_{\Sigma}=0.
\label{A8}
\end{equation}
Both conditions are independent of global coordinates $x^{\alpha}$. Whether the second condition is broken,  the spacetime is singular in  $\Sigma$ and this is  associated to  presence of matter in the hypersurface as
\begin{equation}
S_{ab}=-\frac{\epsilon}{8\pi}\left(\left[K_{ab}\right]-\left[K\right]h_{ab}\right),\,\,\,K\equiv h^{ab}K_{ab}=n^{\alpha}_{;\alpha}.
\label{A9}
\end{equation}
Thus, the stress-energy tensor is correlated to the \textit{jump} in extrinsic curvature from side of $\Sigma$ to other side. The stress-energy tensor on the hypersurface is defined as \cite{PE}
\begin{equation}
T^{\alpha\beta}_{\Sigma}=\delta(l)S^{ab}e^{\alpha}_{\,a}e^{\beta}_{\,b}.
\end{equation}

\section{The  Shell Black in gravitational contraction}
For a external observer in asymptotic far region, the sight correspond to a thin Shell which contracts close to the gravitational radius. Consequently, the spacetime around of this thin Shell, is  symmetrically spherical and static
\begin{equation}
ds^{2}_{+}=-f(r)dt^{2}+\frac{1}{f(r)}dr^{2}+r^{2}d\theta^{2}+r^{2}\sin^{2} \theta d\phi^{2},
\label{A10}
\end{equation}

	\[f(r)=1-\frac{2M}{r}
\]
and inside of Shell, we have got Minkowski-like spacetime
\begin{equation}
ds^{2}_{-}=-dt^{2}+dr^{2}+r^{2}d\theta^{2}+r^{2}\sin^{2}\theta d\phi^{2}.
\label{A20}
\end{equation}
\begin{figure}[h]
\centering
	\includegraphics[width=0.7 \textwidth]{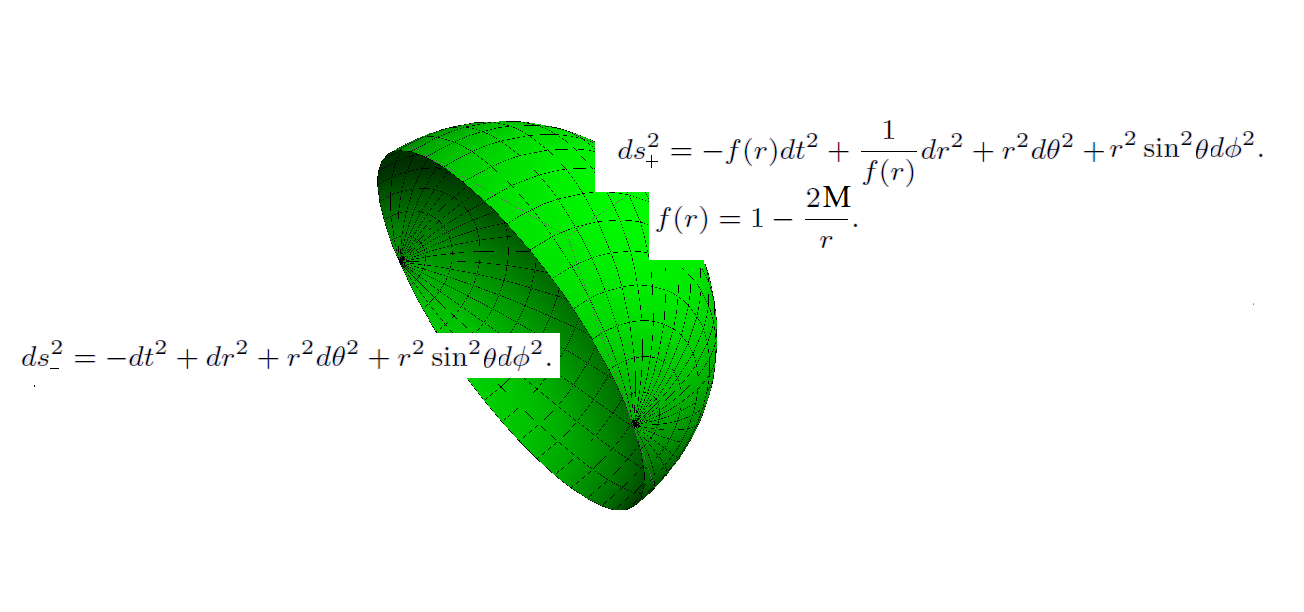}
\caption{A thin Shell which is in a contraction close to the gravitational radius.}
\label{F1}
\end{figure}
%


\subsection{Internal solution $M^{-}$, $u^{\alpha}_{-}$,  $n^{-}_{\alpha}$}
The internal solution is flat, so we can choose the coordinates of $M^{-}$ as
\begin{equation}
t=\bar{T}(\tau),\,\,\,r=R(\tau),
\label{AW31}
\end{equation}
thus, we  have that (\ref{A20}) could be rewritten in terms of (\ref{AW31}) as
\begin{equation}
ds^{2}_{-}=-\left[\dot{\bar{T}}^{2}(\tau)-\dot{R}^{2}(\tau)\right]d\tau^{2}+R^{2}(\tau)d\Omega^{2}.
\label{AW35}
\end{equation}
On other hand, if we can choose the global  coordinates as $x^{\alpha}_{-}=\left(\bar{T}(\tau),R(\tau),\theta,\phi,\right)$ and induced coordinates on $\Sigma$ as $y^{a}=\left(\tau,\theta,\phi\right)$, then, we find that 4-velocity is
\begin{equation}
u^{\alpha}_{-}=e^{\alpha}_{\tau}=\frac{\partial x^{\alpha}_{-}}{\partial \tau}=\left(\dot{\bar{T}}(\tau),\dot{R}(\tau),0,0\right).
\label{AS40}
\end{equation} 
Besides, we have the orthogonality conditions $n^{-}_{\alpha}u^{\alpha}_{-}=0$ and normalization conditions $n^{-}_{\alpha}n^{\alpha}_{-}=1$
\begin{equation}
n^{-}_{\alpha}=\left(-\dot{R}(\tau),\dot{\bar{T}}(\tau),0,0\right).
\label{A451}
\end{equation}
%
\subsection{External solution $M^{+}$, $u^{\alpha}_{+}$,  $n^{+}_{\alpha}$}
The  external solution is a Schwarzschild spacetime. Thus, we can  choose the coordinates of $M^{+}$ as
\begin{equation}
t=T(\tau),\,\,\,r=R(\tau)\,\,\,\mbox{and}\,\,\, f(r)=F(R)=1-\frac{2M}{R}.
\label{AW70}
\end{equation}
So, we have that (\ref{A10}) can be rewritten as
\begin{equation}
ds^{2}_{+}=-\left[FT(\tau)^{2}-\frac{\dot{R}^{2}(\tau)}{F}\right]d\tau^{2}+R^{2}(\tau)d\Omega^{2},
\label{AW71}
\end{equation}
furthermore, the  global coordinates are $x^{\alpha}_{+}=\left(T(\tau),R(\tau),\theta,\phi,\right)$ and induced  coordinates are $y^{a}=\left(\tau,\theta,\phi\right)$. Then, the 4-velocity is
\begin{equation}
u^{\alpha}_{+}=e^{\alpha}_{\tau}=\frac{\partial x^{\alpha}_{+}}{\partial \tau}=\left(\dot{T}(\tau),\dot{R}(\tau),0,0\right).
\label{A40}
\end{equation} 
As a result, we have the orthogonality conditions again $n^{+}_{\alpha}u^{\alpha}_{+}=0$ and normalization conditions $n^{+}_{\alpha}n^{\alpha}_{+}=1$
\begin{equation}
n^{+}_{\alpha}=\left(-\dot{R}(\tau),\dot{T}(\tau),0,0\right)
\label{AW45}
\end{equation}

\subsection{Shell motion equation}
The Shell motion equation is obtained from (\ref{A9})
\begin{equation}
\frac{dR}{d\tau}=\sqrt{\left[\frac{M}{\mu}+\frac{\mu}{2R}\right]^{2}-1},
\end{equation}
Whether we define a new  constant $a$ as the ratio between gravitational mass $M$ and rest mass $\mu$ of Shell, then we get 
\begin{equation}
a=\frac{M}{\mu},\,\,\,\, \frac{dR}{d\tau}=\sqrt{\left[a+\frac{M}{2aR}\right]^{2}-1}.
\label{A210}
\end{equation} 
Such parameter $a$ has got kinetic energy contribution and  gravitational energy potential of particles of Shell. Israel defines $\mu$ is  the nucleonic mass of  particles of Shell, and also he affirms that the mass gravitational contains all energy forms: kinetic, gravitational and internal interaction among these  particles (For this  study is null) \cite{IW2}.

\subsection{Conditions of Shell motion  equation}
It is possible to set up some conditions on kinematic of Shell from (\ref{A210}).That's because for this equation the root cannot be negative, so, we have that
	\[\left[a+\frac{\mu}{2aR}\right]^{2}-1\geq 0.
\]
And, if we  consider the upper limit $R=R_{\mbox{max}}$,  we have \cite{IW2}
	\begin{equation}
	R_{max}=\frac{M}{2a(1-a)}
	\end{equation}
	At this point, is very import to take the  conditions proposed by Israel about Shell motion  equation. these are the following \cite{IW2}
\begin{enumerate}
	\item If the positive binding energy is $a=\cos \alpha <1$ and it is $E=(1-a)\mu<1$. Consequently Shell contracts from resting state with a finity radius $\dot{R}|_{R_{\mbox{max}}}=0$.

	\item If $a=\cosh \alpha >1$,  we obtain a negative binding energy. Thus, the Shell is impelled from infinity with initial velocity null $\dot{R}|_{R\longrightarrow \infty}=0$.
	
	\item When $a=1$,	the binding energy is zero, thus the internal pressure in the Shell also in zero. Thus, the Shell is a thin spherical  layer of dust and this goes down from rest the resting state at $\dot{R}|_{R\longrightarrow \infty}=0$.
	\item In particular case that   $a\longrightarrow\infty$ then $\mu\longrightarrow a$ and therefore  $M=a\mu$, the result  is a Shell made of photons.
\end{enumerate}
	
\subsection{Israel's solution to Shell motion equation}
Being the Shell motion equation already established (\ref{A210}) there  is an interest in studying the their case in the previous section. For this reason Israel presents the following set of  solutions of the Shell motion equation \cite{IW2}
\begin{equation}
r(\lambda)=M\left(\lambda^{2}-\frac{1}{4}\right),
\label{AB220}
\end{equation}
\begin{equation}
t(\lambda)=M\left(\frac{2}{3}\lambda^{3}+\frac{5}{2}\lambda\right)+2M\ln\left| \frac{\lambda-3/2}{\lambda+3/2}\right|+C
\label{AB230},
\end{equation}
where $\lambda$ in (\ref{AB220}) and (\ref{AB230}) is a  free parameter to the solution of (\ref{A210}), and $C$  is a  constant that depends on the initial  conditions in the  kinematic of Shell. Is possible eliminate $\lambda$ in (\ref{AB220}) and (\ref{AB230}), so we obtain
\begin{equation}
t(r)=-M\left[\frac{2}{3}\left(\frac{r}{M}+\frac{1}{4}\right)^{3/2}+\frac{5}{2}\left(\frac{r}{M}+\frac{1}{4}\right)^{1/2}\right]+
2M\ln \left| \frac{\sqrt{\frac{r}{M}+\frac{1}{4}}+3/2}{\sqrt{\frac{r}{M}+\frac{1}{4}}-3/2}\right|+C.
\label{AB235}
\end{equation}
\subsection{ Arenas-Castro solution to Shell motion equation}
\begin{figure}[h]
\centering
	\includegraphics[width=0.7 \textwidth]{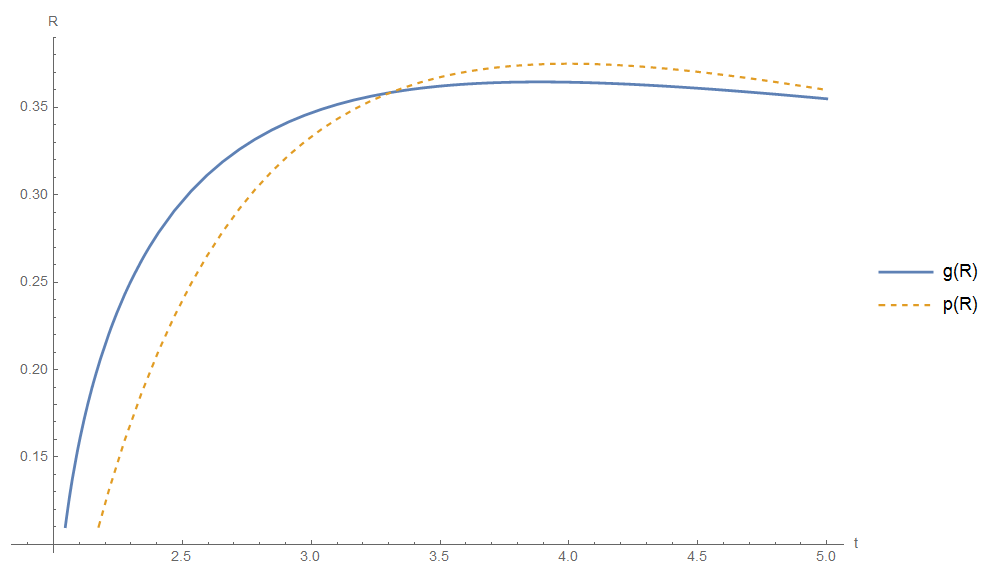}
\caption{The behavior of $g(R)$ and $p(R)$.}
\label{FA10}
\end{figure}

Arenas-Castro  \cite{JR} presents an approximate solution to the Shell motion equation (\ref{A210}). Given the relation between the coordinate time $t$ and proper time  $\tau$ \cite{LL} as
\begin{equation}
d\tau=\sqrt{f(R)}dt,
\end{equation}
where we assume the conditions of case 3, so we write the Shell motion equation as
	\[\frac{1}{\sqrt{f(R)}}\frac{dR}{dt}=-\sqrt{\left[1+\frac{M}{2R}\right]^{2}-1},\,\,\,f(R)=1-\frac{2M}{R}.
\]
\begin{equation}
\frac{dR}{d\tau}=-\sqrt{\left(1-\frac{2M}{R}\right)\left(\frac{M}{4R}+1\right)\frac{M}{R}}.
\label{AB240}
\end{equation}
we define a  function $g(R)$ as
		\begin{equation}
	g(R)=\sqrt{\left(1-\frac{2M}{R}\right)\left(\frac{M}{4R}+1\right)\frac{M}{R}}.
	\label{AB250}
	\end{equation}
Then, for $g(R)$ defined in (\ref{AB250}) another function $p(R)$ can be approximated to  as
\begin{equation} 
p(R)\approx K\left(\frac{M}{R}\right)\left(1-\frac{2M}{R}\right),\,\,\,K\approx 3.
\label{AB260}
\end{equation}	
In the Figure \ref{FA10}, we can note the behavior of $g(R)$ and $p(R)$. We choose $K$ under the following condition
	\[\frac{dg(R)}{dr}|_{R=R_{max}}=0
	\]
	 
%
So, we can rewrite the Shell motion equation as
\begin{equation}
\frac{dR}{dt}=-K\left(\frac{M}{R}\right)\left(1-\frac{2M}{R}\right).
\label{AB270}
\end{equation}		
The integration of (\ref{AB270}) is
	\[\int^{r}_{r_{0}}\frac{dR}{\left(\frac{M}{R}\right)\left(1-\frac{2M}{R}\right)}=-3\int^{t_{s}}_{t_{0}}dt,
	\]
		\begin{equation}
\frac{-6M(t_{s}-t_{0})-4M(r-r_{0})+(r^{2}-r_{0}^{2})}{8M^{2}}=\ln\left|\frac{r-2M}{r_{0}-2M}\right|,
\label{AB280}
\end{equation}
where $r_{s}=2M$ is the Schwarzschild radius,  $\Delta t=t_{s}-t_{0}$ and $\bar{\tau}=\frac{4}{3}M$. Then, the equation (\ref{AB280}) reduces to
\begin{equation}
r-r_{s}=(r_{0}-r_{s})e^{-\frac{\Delta t}{\bar{\tau}}}e^{-\left[\frac{(r-r_{0})}{r_{s}}+\frac{(r^{2}-r_{0}^{2})}{2r_{s}^{2}}\right]}.
\label{AB290}
\end{equation}
It is possible define a new function $\Bar{F}$ in the equation (\ref{AB290})
\begin{equation}
\bar{F}=e^{-\left[\frac{(r-r_{0})}{r_{s}}+\frac{(r^{2}-r_{0}^{2})}{2r_{s}^{2}}\right]},
\end{equation}
the  function  $\Bar{F}$ is expanded in a Taylor series as		$e^{x}=\sum_{n=0}^{\infty}\frac{x^{n}}{n!}$ and it has  an approximation to  Zero order. Thus, $\bar{F}\approx1$	 and   (\ref{AB290})  reduces to
\begin{equation}
r(t)=r_{s}+\underbrace{(r_{0}-r_{s})}_{\Delta r}e^{-\frac{\Delta t}{\bar{\tau}}}= r_{s}+\Delta r e^{-\frac{\Delta t}{\bar{\tau}}}.
\label{AB300}
\end{equation}
We rewrite the equation (\ref{AB300}) as
\begin{equation}
t(r)=\bar{\tau}\ln\left|\frac{r_{0}-r_{s}}{r-r_{s}}\right|,\,\,\,\bar{\tau}=\frac{4}{3}M
\label{AB305}
\end{equation}	
\subsection{ Akhmedov-Godazgar-Popov solution to the Shell motion equation}
\begin{figure}[h]
\centering
	\includegraphics[width=0.7\textwidth]{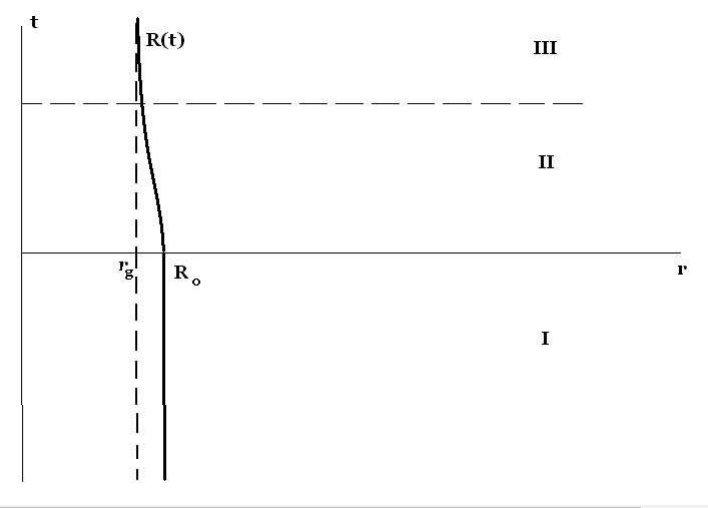}
\caption{The movement of Shell that is seen by an external observer in a asymptotic far region. The  Figure is taken from  \cite{AE}.}
\label{FA20}
\end{figure}

Recently, Akhmedov-Godazgar-Popov \cite{AE} presented a new Shell motion equation.  In this paper, this equation is interpreted as  a new condition of the Shell evolution, additional to Israel's conditions. As a result, the movement of Shell has these stages
\begin{itemize}
	\item The Shell is remained fixed at some  $r(t)=r_{0}$ by an additional force (see Figure \ref{FA20}).
	\item This phase is a highly tuned stage of collapse if one wishes to respect spherical symmetry.
	\item The final phase describes stage where the Shell collapses.
\end{itemize}
%
Akhmedov-Godazgar-Popov obtained the Shell motion equation from the first condition of junction, (\ref{A7}). Thus, they found that
\begin{equation}
\dot{\bar{T}}^{2}-\dot{R}^{2}=F\dot{T}^{2}-\frac{\dot{R}^{2}}{F}.
\label{AB310}
\end{equation}
Nevertheless, the normal vectors $n_{\alpha}^{+}$ and $n_{\alpha}^{-}$   to $\Sigma$ are defined  in (\ref{A451}) and (\ref{AW45}). This vectors obey the orthonormalization condition (\ref{A3}). So, we can take equation (\ref{AB310}) write and find
\begin{equation}
\dot{\bar{T}}^{2}-\dot{R}^{2}=F\dot{T}^{2}-\frac{\dot{R}^{2}}{F}=1.
\label{AB315}
\end{equation}
In the light of this, Akhmedov-Godazgar-Popov discovered a new Shell motion equation which was
	\[\left(\frac{dt_{-}}{d\tau}\right)^{2}-\left(\frac{dR}{d\tau}\right)^{2}=\left(1-\frac{2M}{R}\right)\left(\frac{dt_{+}}{d\tau}\right)^{2}-\frac{1}{\left(1-\frac{2M}{R}\right)}\left(\frac{dR}{d\tau}\right)^{2}=1,
\]
\begin{equation}
\left(1-\frac{2M}{R}\right)\left(\frac{dt_{+}}{d\tau}\right)^{2}-\frac{1}{\left(1-\frac{2M}{R}\right)}\left(\frac{dR}{d\tau}\right)^{2}=1.
\label{AB325}
\end{equation}
To determine solution for the new Shell motion equation, they proposed the following set of conditions on Shell motion

			\begin{enumerate}
				\item The Shell  velocity  measured by comovil observer when $\frac{dR}{d\tau}\neq 0$ is not null, non zero, around the gravitational radius.
				\item Also,  the relation between the coordinate  times $t_{+}$ and  $t_{-}$ is
				\begin{equation}
				dt_{-}=\sqrt{f(r_{s})}dt_{+},\,\,\,f(r_{s})=1-\frac{2M}{r_{s}},\,\,\,\mbox{when}\,\,\,t_{+}\leq 0,
				\end{equation}
				then, this quantity diverges when
				\begin{equation}
				\lim_{t_{+}\to\infty}\frac{dt_{+}}{d\tau}=\infty.
				\end{equation}
				\item Additionally $\left|R-r_{s}\right|\ll r_{s}$.
			\end{enumerate}
			
On  the condition that right handside in comparison to the left handside is neglect in the new Shell motion  equation, (\ref{AB325}),then it's found  that
\begin{equation}
\left(1-\frac{2M}{R}\right)\left(\frac{dt_{+}}{d\tau}\right)^{2}-\frac{1}{\left(1-\frac{2M}{R}\right)}\left(\frac{dR}{d\tau}\right)^{2}\approx 0.
\label{AB335}
\end{equation}		
The integration of (\ref{AB335}) is
	\[-\int^{t_{s}}_{t_{0}}dt_{+}=\int^{r}_{r_{0}}\frac{1}{1-\frac{2M}{R}}dR
\]
\begin{equation}
-(t_{s}-t_{0})=r-r_{0}+2M\ln \left|\frac{r-2M}{r_{0}-2M}\right|
\label{AB340}
\end{equation}		
after short simplification, they found
\begin{equation}
r-2M=(r_{0}-2M)e^{-(r-r_{0})/2M}e^{-\Delta t/2M},\,\,\,\Delta t=t_{s}-t_{0}.
\label{AB350}
\end{equation}
Once again, we can specify a new  function $\bar{F}=e^{-(r-r_{0})/2M}\approx 1$. Then, the solution to the Shell motion is
\begin{equation}
r(t)=r_{s}+\Delta r e^{-\Delta t/2M},\,\,\,r_{s}=2M.
\label{AB360}
\end{equation}
We rewrite the equation (\ref{AB360}) as
\begin{equation}
t(r)=r_{s}\ln\left|\frac{r_{0}-r_{s}}{r-r_{s}}\right|.
\label{AB370}
\end{equation}
\subsection{Comparison between the solutions to the Shell motion equation}
In this section, the comparison among the several solutions that we review previously, is presented. The solutions already taken into account have been considered as the solutions for the Shell motion equation, according to an external observer's sight.

\begin{figure}[h]
\centering
	\includegraphics[width=0.7\textwidth]{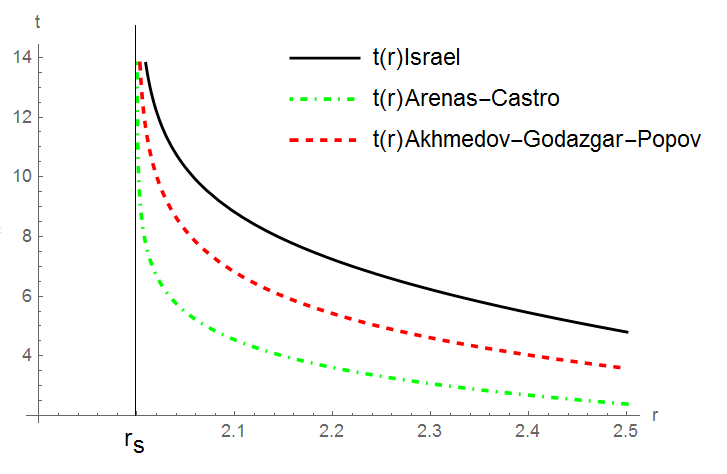}
\caption{Comparison of several solutions of the Shell motion equation close to Schwarzschild radius. The  axes are in cartesian coordinates.}
\label{FA30}
\end{figure}

In the Figure \ref{FA30}, we can note this   solutions comparison. The  $t$ and $x$ axes are in cartesian coordinates. When the Shell is close to Schwarzschild radius, all solutions converge to same behavior.

%

\begin{figure}[h]
\centering
	\includegraphics[width=0.7\textwidth]{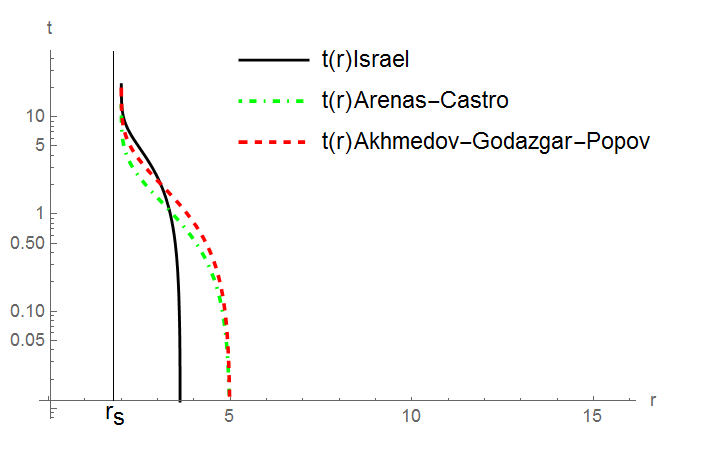}
\caption{Comparison of the several solutions of the Shell motion equation close to Schwarzschild radius. The  $t$ axis is in logarithmic scale \cite{AE}.}
\label{FA40}
\end{figure}

Once more, in the Figure \ref{FA40} the behavior of the solutions studied is shown, where the $t$ axis is in logarithmic scale \cite{AE}. In addition, we can see the same  behavior for the solutions analyzed, when the Shell is close to horizon.

%

Moreover, the Arenas-Castro  (\ref{AB300})  and  Akhmedov-Godazgar-Popov (\ref{AB360}) solutions can rewritten in turtle coordinates	\cite{WR}, $r^{*}=r+2M\ln\left|\frac{r}{2M}-1\right|$. So we find that (\ref{AB300}) reduces to
\begin{equation}
r^{*}_{\mbox{Arenas-Castro}}(t)=r^{*}_{0}-\frac{3}{2}t+\left(r_{s}-r_{0}\right)\left[1-e^{-t/\bar{\tau}}\right],\,\,\,r^{*}_{0}=r_{0}+r_{s}\ln\left|\frac{r_{0}}{r_{s}}-1\right|\,\,\,\mbox{y}\,\,\,\bar{\tau}=\frac{4}{3}M
\label{AB380}
\end{equation}
and also	(\ref{AB360})
\begin{equation}
r^{*}_{\mbox{Akhmedov-Godazgar-Popov}}(t)=r^{*}_{0}-t+\left(r_{s}-r_{0}\right)\left[1-e^{-t/\bar{2M}}\right],\,\,\,r^{*}_{0}=r_{0}+r_{s}\ln\left|\frac{r_{0}}{r_{s}}-1\right|.
\label{ABB390}
\end{equation}

\begin{figure}[h]
\centering
	\includegraphics[width=0.7\textwidth]{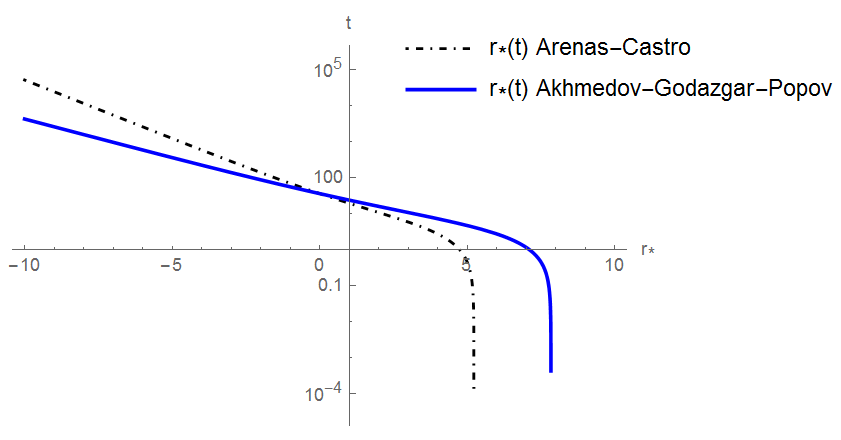}
\caption{The Arenas-Castro and Akhmedov-Godazgar-Popov solutions in turtle  coordinates.}
\label{FA50}
\end{figure}

In the Figure \ref{FA50}, we observe the Arenas-Castro and Akhmedov-Godazgar-Popov solutions in turtle  coordinates.  This graph allows us to see that $\Sigma$ is  timelike  hypersurface , which turns to be null hypersurface, when $r^{*}_{i}(t)\longrightarrow r_{s}$ \cite{AE}
%
\subsection{Behavior of the Shell near to horizon}	
We take the Arenas-Castro solution for Shell movement in (\ref{AB300})
	\begin{equation}
r(t)= r_{s}+\Delta r e^{-\frac{\Delta t}{\bar{\tau}}},\,\,\,\bar{\tau}=\frac{4}{3}M,
\label{AB390}
\end{equation}
where $t$ is the coordinate time for an external observer.
\\
The Shell movement can be parametrized as
\begin{equation}
r(t)=nr_{s},\,\,\,r_{s}=2M.
\label{AB400}
\end{equation} 
As a result, the Shell goes from a distance leaving the resting state $n$ times the gravitational radius, and it stops when is near by the horizon
	\begin{equation}
r(t_{0})=r_{0}=2Mnr_{s},\,\,\,r(t_{1})=r_{s},\,\,\,t_{1}\gg t_{0}.
\label{AB410}
\end{equation}	
In the same way, the coordinate time can be expressed as $n$ times the Planck time as
\begin{equation}
t=nt_{p}.
\label{AB420}
\end{equation}	
Also, we have that $\Delta r$ is rewritten as (see next section about this equation)
\begin{equation}
\epsilon=2\Delta r=2nl_{p}.
\label{AB430}
\end{equation}
Replacing the (\ref{AB410})-(\ref{AB430})  equations in (\ref{AB390}) it is obtained
\begin{equation}
n=\frac{1}{2M}\left[2M+n e^{-\frac{3n}{4M}}\right].
\label{AB440}
\end{equation}	
The equation (\ref{AB440}) is solved when the  Shell mass is know (we try several parametric  form to the (\ref{AB420}) and (\ref{AB430}) equations, having the same result.)

		\begin{table}			
\begin{center}
	\centering
  \begin{tabular}{ | l | c |  }
    \hline
	$M_{\odot}$&n\\   \hline
			$10^{0}$&$1.24469$\\   \hline
				$10^{1}$&$1.04846$\\   \hline
					$10^{2}$&$1.00499$\\   \hline
						$10^{3}$&$1.00050$\\   \hline
							$10^{4}$&$1.00005$\\   \hline
								$10^{5}$&$1.00000$\\   \hline
									$10^{6}$&$1.00000$\\   \hline
	
  \end{tabular}
	\caption{Numerical solution of the equation (\ref{AB440}) for different solar mass $M_{\odot}$.}
	\label{ABB1}
\end{center}
\end{table}
In the chart \ref{ABB1}, it is observed that the value $n=1$ is independent of  solar mass of Shell. As a consequence we fin it from (\ref{AB430}) that				
\begin{equation}
\epsilon=2l_{p},
\label{AB450}
\end{equation}
as the minimum value that $\epsilon$ can taken near by the horizon. This is import, because for  an external observer is able to see that the Shell movement is stops near to  the Schwarzschild radius and never goes beyond from this point. 
\\
In the Figures \ref{FA60} to \ref{FA80}, the Shell's kinematic  is presented   according  to the Arenas-Castro solution. So, for an external observer, the Shell movement is stopped, its 3-velocity and 3-acceleration are null close to the gravitational radius. In such conditions,      any effect can be attributed to the presence of the horizon, due to the fact that the inner region $r<2M$ far beyond is completely unknown for any external observer. Also, the minimal distance between the horizon and the Shell is $\epsilon=2l_{p}$ for finite time equal to universe age. Just when
\begin{equation}
\lim_{t\to\infty}r(t)=r_{s}.
\end{equation}

\subsection{Thermodynamics properties of the Dirac's field near to gravitational radius.}
The spacetime around the Shell in gravitational contraction is 
\begin{equation}
ds^{2}=-f(r)dt^{2}+\frac{1}{f(r)}dr^{2}+r^{2}d\theta^{2}+r^{2}\sin^{2}d\phi^{2},\,\,\, f(r)=1-\frac{2M}{r}.
\end{equation}
The Shell movement, here, is described by the Arenas-Castro solution \cite{JR}
\begin{equation}
r(t)=r_{s}+\Delta r e^{-t/ \bar{\tau}},\,\,\,\, \Delta r=r_{0}-r_{s}, \,\,\,\,r_{s}=\frac{2GM}{c^{2}},\,\,\,\, \bar{\tau}=\frac{4GM}{3 c^{3}}.
\label{B1}
\end{equation}
Furthermore, for great values of the coordinate time we have that
	\[\lim_{t\longrightarrow t_{p}}r(t)=r_{s}.
\]
Therefore, we calculate the differential from equation (\ref{B1}) and the result is
\begin{equation}
\frac{dr}{dt}=-\frac{\Delta r e^{-t/\bar{\tau}}}{\bar{\tau}}.
\label{B10}
\end{equation}
Where the $(-)$ in the equation (\ref{B10}), shows that  the ratio is decreasing  related to the coordinate time.

\begin{figure}[h]
\centering
	\includegraphics[width=0.7\textwidth]{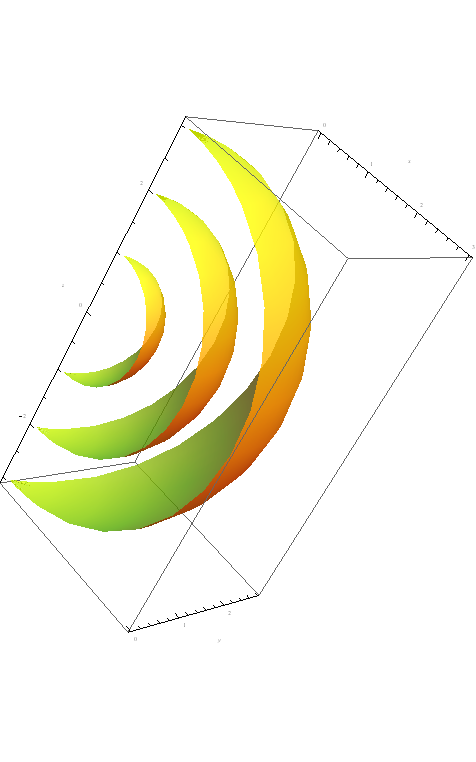}
\caption{The Shell is in gravitational contraction from infinity to near the horizon.}
\label{}
\end{figure}
The entropy of the Dirac field close  to horizon under high energies approximation  \cite{MS, RW3} corresponds to a description of standard statistical physics. Also, particles with rest-mass $m$, energy $E$, 3-momentum $p$ and 3-velocity, that are measures for an external observer. So, the density energy $\rho$, pressure $P$ and entropy density $s(r)$ are
\begin{equation}
\rho=-\int^{\infty}_{0}\frac{E}{e^{E/T(r)}+1}\frac{4\pi^{2}}{h^{3}},
\end{equation}
\begin{equation}
p=\frac{1}{3}\int^{\infty}_{0}\frac{pv}{e^{E/T(r)}+1}\frac{4\pi^{2}}{h^{3}}dp,
\end{equation}
\begin{equation}
S=\int_{Vol}s\, dV,\,\,\,s=\beta(\rho+P)=\frac{4NT^{3}}{\pi^{2}},
\label{B30}
\end{equation}
where $N$ is quantum fields number and their helicities, the  local temperature $T(r)$ which is given by Tolman's law \cite{TR}
\begin{equation}
T(r)=\frac{T_{\infty}}{\sqrt{f(r)}}.
\end{equation}
\\
Thus, 3-volume differential element $dV$, is determinated by
\begin{equation}
dV=\frac{4\pi}{\sqrt{f(r)}}r^{2}dr=\frac{4\pi}{\sqrt{f(r)}}\left(r_{s}+\Delta r e^{-t/ \bar{\tau}}\right)^{2}\frac{\Delta r e^{-t/\bar{\tau}}}{\bar{\tau}}dt.
\label{B40}
\end{equation}
With an approximation to the first order, we have than $\Delta r \gg \Delta r^{2},\Delta r^{3}$. Therefore, the 3-volume differential element reduces to
\begin{equation}
dV \approx \frac{4\pi}{\sqrt{f(r)}}\left(\frac{r_{s}^{2}\Delta r e^{-t/\bar{\tau}}}{\bar{\tau}}\right)dt.
\label{BB50}
\end{equation}
\\
So the equation (\ref{B30}) is rewritten follows
\begin{equation}
S=\frac{4N}{\pi^{2}}T_{\infty}^{3}\frac{4\pi r_{s}^{2}}{\bar{\tau}}\Delta r \int \frac{e^{t/\bar{\tau}}}{f^{2}(r)}dt,
\label{B50}
\end{equation}
where, the horizon area is $A_{\mathcal{H}}=4\pi r_{s}^{2}$ and $f(r)=f(r(t))$. Near to the  Schwarzschild radius, we  have that this can be reexpressed as
\begin{equation}
f(r(t))\approx 2 \kappa_{0} (r(t)-r_{s})=2 \kappa_{0} \Delta r e^{-t/\bar{\tau}},
\label{B60}
\end{equation}
where the surface gravity on the horizon is \cite{MS}
\begin{equation}
\kappa_{0}=-\frac{1}{2}\frac{df(r)}{dr}|_{r=r_{s}}=\frac{1}{4M}.
\end{equation}
After short simplification in the equation (\ref{B50}), we find
\begin{equation}
S=\frac{4N}{\pi^{2}}\left(\frac{T_{\infty}^{3}}{\kappa_{0}^{2}}\right) \frac{A_{\mathcal{H}}}{4}\frac{1}{\Delta r}\left[\frac{1}{\bar{\tau}}\int^{t_{u}}_{0} e^{t/\bar{\tau}}dt\right],
\label{B70}
\end{equation}
where the integral in the equation (\ref{B70}) corresponds to a dimensionless constant depending on the coordinate time  as
\begin{equation}
C_{1}=\frac{1}{\bar{\tau}}\int^{t_{u}}_{0} e^{t/\bar{\tau}}dt.
\label{B80}
\end{equation}
Thus, the equation (\ref{B70}) according to (\ref{B80}), is reduces to
\begin{equation}
S=\frac{4N}{\pi^{2}}\left(\frac{T_{\infty}^{3}}{\kappa_{0}^{2}}\right) \frac{A_{\mathcal{H}}}{4}\frac{1}{\Delta r}C_{1}.
\label{B90}
\end{equation}
On the other hand, the proper altitude  above the horizon $\alpha$ is \cite{LL}
\begin{equation}
\alpha=\int^{r_{1}}_{r_{0}}\frac{dr}{\sqrt{f(r)}}=\frac{1}{\bar{\tau}}\int^{t_{u}}_{0}\frac{\Delta re^{-t/\bar{\tau}}}{\sqrt{2 \kappa_{0} \Delta r e^{-t/\bar{\tau}}}}dt=\sqrt{\frac{\Delta r}{2\kappa_{0}}} \left[\frac{1}{\bar{\tau}} \int^{t_{u}}_{0} e^{-t/\bar{\tau}} dt \right].
\label{B100}
\end{equation}
The expossed previously is rewritten in terms of a additional constant as
 \begin{equation}
\alpha=\sqrt{\frac{\Delta r}{2\kappa_{0}}}C_{2},
\label{B110}
\end{equation}

 \begin{equation}
C_{2}=\frac{1}{\bar{\tau}} \int^{t_{u}}_{0} e^{-t/\bar{\tau}} dt.
\label{B115}
\end{equation}
hence, the Dirac field entropy,  (\ref{B90}), near by to horizon is reduced to
\begin{equation}
S_{\mbox{Shell}}=\left[\frac{N}{90\pi \alpha^{2}}\left(\frac{T_{\infty}}{\kappa_{0}/2\pi}\right)^{3}\frac{A}{4}\right]\frac{C_{1}C^{2}_{2}}{4}=\frac{S_{\mbox{BWM}}}{4}C_{1}C^{2}_{2},
\label{B120}
\end{equation}
where $S_{\mbox{BMW}}$ is the entropy for quantum fields, and  it works according to Mukohyama-Israel \cite{MS}. 
Also, whether   $S_{\mbox{BWM}}=S_{\mbox{Shell}}$ and $\Delta r_{\mbox{BWM}}=\Delta r_{\mbox{Shell}}$, we can have  the  values for the  constants, so
\begin{equation}
C_{1}=1,\,\,\,C_{2}=\pm 2.
\end{equation}
And the Dirac field entropy is
\begin{equation}
\Delta r_{\mbox{Shell}}=\frac{4\Delta r_{\mbox{BWM}}}{C^{2}_{2}},\,\,\,\Delta r_{\mbox{BWM}}=\frac{1}{2}\kappa_{0} \alpha^{2}. 
\label{B130}
\end{equation}
\begin{equation}
S_{\mbox{Shell}}=\frac{N}{90\pi \alpha^{2}}\left(\frac{T_{}\infty}{\kappa_{0}/2\pi}\right)^{3}\frac{A}{4},\,\,\, \Delta r_{\mbox{Shell}}=\frac{\kappa_{0} \alpha^{2}}{2}.
\label{B140}
\end{equation}
 An external observer can measure that the temperature of the Dirac field is  equal to Hawking temperature, $T_{\infty}=T_{H}=\frac{\kappa_{0}}{2\pi}$ , so the the Dirac field  entropy is
\begin{equation}
S_{\mbox{Shell}}=S_{BH}=\frac{1}{4}\frac{A}{l_{p}^{2}},
\end{equation}
and the proper altitude  above the horizon,  $\alpha$ \cite{MS} is
\begin{equation}
\alpha=l_{P}\sqrt{\frac{C_{1}C^{2}_{2}N}{360\pi}}=l_{P}\sqrt{\frac{N}{90\pi}}.
\label{B150}
\end{equation}

\subsection{About the values of the constants $C_{1}$ and $C_{2}$}
The equation (\ref{B80}) is rewritten as
	\[\frac{dC_{1}}{dt}=\frac{1}{\bar{\tau}}e^{t/\bar{\tau}},
	\]

	\[ \frac{dC_{1}}{dr} \frac{dr}{dt}=\frac{1}{\bar{\tau}}e^{t/\bar{\tau}},
\]
where $dr/dt$ is given by (\ref{B1})  and the  times coordinate by (\ref{AB305}). Then, $C_{1}$ is reduced
\begin{equation}
C_{1}=-\Delta r \int_{r_{0}}^{r_{s}+\epsilon}\frac{1}{(r-r_{s})^{2}}dr=\frac{r_{0}-r_{s}-\epsilon}{\epsilon}=\frac{\Delta r}{\epsilon}-1.
\label{B160}
\end{equation}
The physical analysis of (\ref{B160}) is  complex, since their values can be in a different order of magnitude
	\[r_{0}\sim 10^{23}m,\,\,\,r_{s}\sim 10^{3}m\,\,\, y\,\,\,\epsilon\sim 10^{-35}m, 
\]
All the above, considering a black hole of one solar mass, $m_{\odot}\sim 10^{30}kg$. So, under these  conditions, a different approach should be considered, as the values of $r$ multiples of Schwarzschild radius, $r_{s}$. Therefore, we can write
	\[r_{0}=r_{s}*10^{m},\,\,\,\epsilon=r_{s}*10^{-n},
\]
Thus, we have that the initial radius $r_{0}$ and the cutoff $\epsilon$ are multiples from gravitational radius $r_{s}$. Thus $m$ and $n$ are not necessarily integer

	\[-\infty\leq m,n\leq \infty.
\]
In the special case that $m=n=0$, we find
\begin{equation}
C_{1}=\frac{10^{m}-1-10^{-n}}{10^{-n}}=10^{m+n}-10^{-n}-1,
\label{B170}
\end{equation}

\begin{equation}
C_{1}=10^{2m}-10^{-m}-1=1.
\label{B180}
\end{equation}
This coincides with Mukohyama- Israel method \cite{MS}, owing to the minimum value that takes $C_{1}$ close to horizon, that is
\begin{equation}
C_{1}(r_{s}+\epsilon)=1,
\label{B190}
\end{equation}
also the ratio between $C_{1}$ and the coordinate time  $t$, is an increasing function
\begin{equation}
\frac{dC_{1}}{dt}=\frac{1}{\bar{\tau}}e^{t/\bar{\tau}}.
\end{equation}
For the  value of $C_{2}$, from the equation (\ref{B115}), can be rewritten as
\begin{equation}
\frac{dC_{2}}{dt}=\frac{1}{\bar{\tau}}e^{-t/\bar{\tau}},
\label{B200}
\end{equation}

where (\ref{B200}) has  a boundary condition $C_{2}(r_{s}+\epsilon)=2$, so we have 
\begin{equation}
\frac{dC_{2}}{dr}\left(\frac{dr}{dt}\right)=\frac{1}{\bar{\tau}}e^{-t/\bar{\tau}},
\label{B210}
\end{equation}

also, (\ref{B210}) is reduced to
\begin{equation}
\frac{dC_{2}}{dr}=-\frac{1}{\Delta r}.
\label{B220}
\end{equation}
And integrating the equation (\ref{B220}), we have
\begin{align}
C_{2}&=-\frac{1}{\Delta r}\int^{r_{s}+\epsilon}_{r_{0}}dr
 	\notag\\
		&\hspace{0.0cm}
=\frac{1}{\Delta r}\left[\Delta r-\epsilon\right]=1-\frac{\epsilon}{\Delta r}=1-\frac{\epsilon}{r_{0}-r_{s}}
	\notag\\
		&\hspace{0.0cm}
\propto \frac{1}{r_{0}}.
\label{B230}
\end{align}
This implies that, $ C_ {2} $ is an decreasing function, where under the boundary condition $C_{2}(r_{s}+\epsilon)=2$. So that, $\epsilon$ is reduces to
\begin{equation}
\epsilon=-\Delta r.
\end{equation}

Nevertheless, whether we consider the ratio $\frac{C_{1}}{C_{2}}$ as
\begin{equation}
\frac{C_{1}}{C_{2}}=\frac{\Delta r}{\epsilon},
\end{equation}
\begin{equation}
\epsilon=\frac{C_{2}}{C_{1}}\Delta r.
\end{equation}
Then is possible define a new  constant $n|_{r_{s}+\epsilon}$ as
\begin{equation}
n|_{r_{s}+\epsilon}=\left(\frac{C_{2}}{C_{1}}\right)_{r_{s}+\epsilon}=2,
\end {equation}
so, the minimum value that $\epsilon$ that is got to the horizon is
\begin{equation}
\epsilon|_{r_{s}+\epsilon}=n \Delta r=2 \Delta r
\label{CCK39}
\end{equation}
Along this analysis, we have to take into account that the constants $C_{i}$ and $\epsilon$ arose from the Shell's kinematic.

Once, we have the Dirac's field  entropy near to horizon, $S_{\mbox{Shell}}$, we can calculate other thermodynamics properties of fermionic field. So, we write from  equation (\ref{B120})
\begin{equation}
S_{\mbox{Shell}}=\frac{N\pi^{2}}{180\alpha^{2}}\left(\frac{T^{3}_{\infty}}{\kappa^{3}_{0}}\right)A\,C_{1}C^{2}_{2}.
\label{B250}
\end{equation}
\begin{itemize}
	\item The Helmholtz free energy, $F$ is written as 
		\[S_{\mbox{Shell}}=-\left(\frac{\partial F}{\partial T_{\infty}}\right)_{V},
	\]
		\begin{equation}
	F=-\frac{N\pi^{2}}{720\alpha^{2}}\left(\frac{T^{4}_{\infty}}{\kappa^{3}_{0}}\right)AC_{1}C^{2}_{2}.
	\label{B260}
	\end{equation}
	
	\item The internal energy, $E$ 
			\[dE=dF+T_{\infty}dS,
	\]
	\begin{equation}
	E=\frac{N\pi^{2}}{240\alpha^{2}}\left(\frac{T^{4}_{\infty}}{\kappa^{3}_{0}}\right)AC_{1}C^{2}_{2}.
	\label{B270}
	\end{equation}
	\item The specific heat capacity at a volume constant  ($C_{V}$)  and  the specific heat capacity at pressure constant ($C_{P}$) are 
	\begin{equation}
	C_{V}=\left(\frac{\partial E}{\partial T_{\infty}}\right)_{V},\,\,\,C_{P}=T_{\infty}\left(\frac{\partial S}{\partial T_{\infty}}\right)_{P},
	\end{equation}
	
also \cite{ZM}
	\begin{equation}
	\lim_{T\to T_{H}}C_{V}=C_{P},\,\,\,T_{H}=\frac{1}{8\pi M_{\odot}} \approx 10^{-8}K.
	\end{equation}
therefore, we find
	\begin{equation}
	C_{V}=C_{P}=\frac{N\pi^{2}}{60\alpha^{2}}\left(\frac{T^{3}_{\infty}}{\kappa^{3}_{0}}\right)AC_{1}C^{2}_{2}.
	\label{B280}
	\end{equation}
	
\item  And  the internal pressure is  calculated ($P$) as
	\[P=-\left(\frac{\partial F}{\partial V}\right)_{T_{\infty}}=-\frac{1}{\alpha}\left(\frac{\partial F}{\partial A}\right)_{T_{\infty}}
\]
As a result we obtain
	\begin{equation}
	P=\frac{N\pi^{2}}{720\alpha^{3}}\left(\frac{T^{3}_{\infty}}{\kappa^{3}_{0}}\right)C_{1}C^{2}_{2}.
	\label{BB290}
	\end{equation}
		\end{itemize}
		
In this point, is  very relevant to mention that if the Shell's entropy is  equal to BWM entropy, we have $S_{\mbox{Shell}}=S_{\mbox{BWM}}$. And  if $C_{1}=1$, $C_{2}=2$, the species number $N=1$ for the  quantum fields, also the proper altitude  above the horizon  $\alpha$ given by (\ref{B150}). Then, the  thermodynamics properties are reduce to		

\begin{equation}
S_{\mbox{Shell}}=\frac{\pi^{2}}{45\alpha^{2}}\left(\frac{T^{3}_{\infty}}{\kappa^{3}_{0}}\right)A.
\label{B290}
\end{equation}
	 	\begin{equation}
	F=-\frac{\pi^{2}}{180\alpha^{2}}\left(\frac{T^{4}_{\infty}}{\kappa^{3}_{0}}\right)A.
	\end{equation}
	\begin{equation}
	E=\frac{\pi^{2}}{60\alpha^{2}}\left(\frac{T^{4}_{\infty}}{\kappa^{3}}_{0}\right)A.
	\label{B300}
	\end{equation}
\begin{equation}
	C_{V}=C_{P}=\frac{\pi^{2}}{15\alpha^{2}}\left(\frac{T^{3}_{\infty}}{\kappa^{3}_{0}}\right)A.
	\end{equation}
\begin{equation}
	P=\frac{\pi^{2}}{180\alpha^{3}}\left(\frac{T^{3}_{\infty}}{\kappa^{3}_{0}}\right).
	\end{equation}
These properties coincides  with  \cite{RW1,RW2}, for the spacetime $3+1$ and  the  spacetime $D+1$ with a  scalar hot quantum field.
\section{The species problem}
Now, we  consider the  species problem for the entanglement entropy, for fermionic  field \cite{RW3} and bosonic field \cite{MS, FD, RW1, RW2, JR2} near by the horizon.

The  fermionic field entropy \cite{RW3}, (\ref{B290}),  is 
\begin{equation}
S_{\mbox{F}}=\frac{\pi^{2}}{45\alpha^{2}}\left(\frac{T^{3}_{\infty}}{\kappa^{3}_{0}}\right)A
\label{PR1}
\end{equation}
and the internal energy of fermionic field  (\ref{B300}) is
		\begin{equation}
	E_{\mbox{F}}=\frac{\pi^{2}}{60\alpha^{2}}\left(\frac{T^{4}_{\infty}}{\kappa_{0}^{3}}\right)A.
	\label{PRR10}
	\end{equation}
	The unification of (\ref{PR1}) and (\ref{PRR10}) with $\beta=\frac{1}{T_{\infty}}$  \cite{CY} is
\begin{equation}
S_{\mbox{F}}=\frac{4}{3}\beta E_{\mbox{F}}.
\label{PR10}
\end{equation}
In the same  way, for bosonic field, we have that the entropy \cite{MS, RW1, RW2, JR2} is
\begin{equation}
S_{\mbox{B}}=\frac{1}{90\pi \alpha^{2}}\left(\frac{T^{3}_{\infty}}{\left(\kappa_{0}/2\pi\right)^{3}}\right)A
\label{PR20}
\end{equation}
And the free energy is
\begin{equation}
S=-\left(\frac{\partial F}{\partial T_{\infty}}\right)_{V},
\label{PR30}
\end{equation}
\begin{equation}
F_{\mbox{B}}=-\frac{\pi^{2}}{180\alpha^{2}} \left(\frac{T^{4}_{\infty}}{\kappa_{0}^{3}}\right)A.
\label{PR40}
\end{equation}
So, the  internal energy $E$ is 
\begin{equation}
dE=dF-T_{\infty}dS
\end{equation}
\begin{equation}
E_{\mbox{B}}=\frac{\pi^{2}}{60\alpha^{2}}\left(\frac{T^{4}_{\infty}}{\kappa_{0}^{3}}\right)A.
\label{PR50}
\end{equation}
This shows that  for the quantum fields studied, we find that
\begin{equation}
E=E_{\mbox{B}}=E_{\mbox{F}}=\frac{\pi^{2}}{60\alpha^{2}}\left(\frac{T^{4}_{\infty}}{\kappa_{0}^{3}}\right)A.
\label{PR60}
\end{equation}
For this reason, the entropy of the bosonic field $S_{\mbox{B}}$ is written as
\begin{equation}
S_{\mbox{B}}=\frac{4}{3}\beta E_{\mbox{B}}.
\label{PR70}
\end{equation}
\\
Whether the quantum fields are in thermal equilibrium, among themselves and the horizon, the product is
	\[ T_{\mbox{B}}=T_{\mbox{F}}=T_{H}.
\]
Also, the  total entropy is the sum of the partial entropies of every field, so we get
\begin{align}
S&=S_{\mbox{B}}+S_{\mbox{F}}
\notag\\
&\hspace{0.0cm}
=\frac{4}{3}\beta E_{\mbox{B}}+\frac{4}{3}\beta E_{\mbox{F}}
\notag\\
&\hspace{0.0cm}
=\frac{4}{3}\beta \left[E_{\mbox{B}}+E_{\mbox{F}}\right]
\notag\\
&\hspace{0.0cm}
=\frac{4}{3}\beta E,\,\,\,\,E=E_{\mbox{B}}+E_{\mbox{F}}.
\label{PR80}
\end{align} 

Moreover, from the equation (\ref{PR60}), the  equation (\ref{PR80}) is rewritten as
\begin{equation}
S=\frac{8}{3}\beta\frac{\pi^{2}}{60\alpha^{2}}\left(\frac{T_{\infty}^{4}}{\kappa_{0}^{3}}\right)A,
\label{PR90}
\end{equation}
where $\beta=1/ \frac{1}{T_{\infty}}$. For the proper altitude  above the horizon, $\alpha$, which was defined previously, (\ref{B150}) 
\begin{equation}
\alpha=l_{P}\sqrt{\frac{N}{90\pi}}.
\label{PR100}
\end{equation}
where $N=2$, for quantum fields, if these fields are in thermal equilibrium with the horizon at a temperature Hawking, $T_{H}=\frac{\kappa_{0}}{2\pi}$. This results in the Bekenstein-Hawking entropy, $S_{BH}$
\begin{equation}
S=\frac{1}{4}\frac{A}{l_{p}^{2}}.
\label{PR110}
\end{equation}
This result is very interesting, due to the Bekenstein-Hawking entropy for  black holes is independent of the  quantum fields near to horizon, their nature, their spins and the internal interaction of quantum fields researched.

The Bekenstein-Hawking entropy arose from thermal atmosphere around the horizon, which is seen  by an external observer in a far asymptotic region.
\section{Summary and Discussion}

Along this paper, we review the Shell equations and their kinematic behavior near by the Schwarzschild radius. Also, we  chose a solution for the Shell motion equation, so it was possible to determine the minimal value of  cutoff close to the horizon for the entanglement entropy for the  Dirac's field
\begin{equation}
\epsilon|_{r_{s}+\epsilon}=n \Delta r=2 \Delta r.
\label{SAD10}
\end{equation}

So that, we found the entanglement entropy for the  Dirac's field near by the gravitational radius as

\begin{equation}
S_{\mbox{Shell}}=S_{\mbox{F}}=\frac{\pi^{2}}{45\alpha^{2}}\left(\frac{T^{3}_{\infty}}{\kappa^{3}_{0}}\right)A,
\label{SAD100}
\end{equation}
where $\alpha$ is the proper altitude above the horizon, according to (\ref{B100}).

With the condition that $T_{\infty}=T_{H}=\frac{\kappa_{0}}{2\pi}$, $S_{\mbox{Shell}}$ coincides with 
 $S_{BH}=\frac{1}{4}\frac{A}{l_{p}}$.

About the species  problem, we showed for a case  of two fields that the entropy is independent of nature and the number of fields near to the horizon: 
\begin{equation}
S_{\mbox{Ent}}=S_{\mbox{B}}+S_{\mbox{F}}=\frac{4}{3}\beta E=\frac{1}{4}\frac{A}{l_{p}^{2}},
\end{equation}
 where $\beta=\frac{1}{T_{\infty}}$, $T_{\infty}=T_{H}$ and $E=E_{\mbox{B}}+E_{\mbox{F}}$. The  Bekenstein-Hawking  entropy, $S_{BH}$ arise  from  thermal atmosphere around to  the horizon.

We consider that this result is  completely generalizable for any number of  fields.
\appendix
\section{The Shell motion equation}
\subsection{Extrinsic curvature seen from $M^{+}$}
Once, the normal vector $n^{\alpha}$ is known, is possible calculate  the extrinsic curvature from (\ref{AW6})
\begin{equation}
K_{\alpha\beta}=n_{\alpha;\beta},
\label{AW55}
\end{equation}
this calculated on external solution, $M^{+}$, then we have
\begin{equation}
K^{+}_{\theta\theta}=\dot{T}RF,\,\,\,K^{+\theta}_{\theta}=g^{\theta\theta}K^{+}_{\theta\theta}=\frac{\beta_{+}}{F},\,\,\,\beta_{+}=\dot{T}F
\label{AW65}
\end{equation}
\begin{equation}
K^{+}_{\phi\phi}=\dot{T}RF\sin^{2}\theta,\,\,\,K^{+\phi}_{\phi}=g^{\phi\phi}K^{+}_{\phi\phi}=\frac{\beta_{+}}{R}
\label{A75}
\end{equation}
and finally
\begin{equation}
K^{+}_{\tau\tau}=-F'\dot{T},\,\,\,K^{+\tau}_{\tau}=g^{\tau\tau}K^{+}_{\tau\tau}=\frac{\dot{\beta}_{+}}{F},\,\,\,\dot{\beta}_{+}=F'\dot{R}\dot{T}.
\label{A85}
\end{equation}
\subsection{Extrinsic curvature seen from $M^{-}$}
As it was developed in the last section, the extrinsic curvature components for  external, $M^{-}$, solution are
\begin{equation}
K^{-}_{\theta\theta}=\dot{T}R,\,\,\,K^{-\theta}_{\theta}=g^{\theta\theta}K^{-}_{\theta\theta}=\frac{\beta_{-}}{R},\,\,\,\beta_{-}=\sqrt{1-\dot{R}^{2}}
\label{A95}
\end{equation}
\begin{equation}
K^{-}_{\phi\phi}=\dot{\dot{T}}R\sin^{2}\theta,\,\,\,K^{-\phi}_{\phi}=g^{\phi\phi}K^{-}_{\phi\phi}=\frac{\beta_{-}}{R}
\label{AW105}
\end{equation}
\begin{equation}
K^{-}_{\tau\tau}=0,\,\,\,K^{-\tau}_{\tau}=g^{\tau\tau}K^{-}_{\tau\tau}=0
\label{A115}
\end{equation}
\subsection{The Shell motion equation}
The Shell motion equation is obtained from the equation (\ref{A9})
\begin{equation}
S^{a}\,_{b}=-\frac{\epsilon}{8\pi}\left(\left[K^{a}\,_{b}\right]-\left[K\right]h^{a}\,_{b}\right),\,\,\,K\equiv h^{ab}K_{ab}=n^{\alpha}_{;\alpha}.
\label{AW120}
\end{equation}
Besides, it is obtained by using  the orthonormal condition of the induced metric, $h^{a}\,_{a}=1$. So from  this equation (\ref{AW120}), we have that
\begin{align}
S^{\tau}\,_{\tau}&=-\frac{\epsilon}{8\pi}\left[ K^{+\tau}\,_{\tau}-	K^{-\tau}\,_{\tau}-\left(K^{+}-K^{-}\right)h^{\tau}\,_{\tau}\right]
\notag\\
			&\hspace{0.0cm}
			=-\frac{\epsilon}{8\pi}\left[ K^{+\tau}\,_{\tau}-	K^{-\tau}\,_{\tau}-K^{+\tau}\,_{\tau}+	K^{-\tau}\,_{\tau}- K^{+\theta}\,_{\theta}+	K^{-\theta}\,_{\theta}-K^{+\phi}\,_{\phi}+K^{-\phi}\,_{\phi}						\right]
			\notag\\
			&\hspace{0.0cm}
			=-\frac{\epsilon}{8\pi} (-2) \left[K^{+\theta}\,_{\theta}-K^{-\phi}\,_{\phi}	\right].
			\label{AW130}
\end{align}
With the  values of $K^{+\theta}\,_{\theta}$ and $K^{-\phi}\,_{\phi}$ defined in (\ref{AW65}) and (\ref{AW105}), the equation (\ref{AW130}) is rewritten as
\begin{equation}
	S^{\tau}\,_{\tau}=-\frac{\epsilon}{8\pi} (-2) \left[\frac{\beta_{+}}{R}-\frac{\beta_{-}}{R}	\right].
\label{AW140}
\end{equation}
\\
The surface tensor $S^{ab}=\sigma u^{a}u^{b}$ is write as $S^{a}\,_{b}=\sigma u^{a}u_{b}$, also with the normalization condition for 4-velocity $u^{a}u_{a}=-1$, we  find
\begin{equation}
S^{a}\,_{a}=\sigma u^{a}u_{a}=-\sigma,\,\,\,S^{\tau}\,_{\tau}=-\sigma. 
\label{A150}
\end{equation}
Replacing (\ref{A150}) into (\ref{AW140}), we obtain 
\begin{equation}
-\sigma=\frac{1}{4\pi} \left[\frac{\beta_{+}}{R}-\frac{\beta_{-}}{R}	\right]
\label{A160}
\end{equation}
in the same way, for the $S^{\theta}\,_{\theta}$ component
\begin{equation}
S^{\theta}\,_{\theta}=-\frac{\epsilon}{8\pi}\left[ K^{+\theta}\,_{\theta}-	K^{-\theta}\,_{\theta}-\left(K^{+}-K^{-}\right)h^{\theta}\,_{\theta}\right]
\end{equation}
After replaced all the values of the extrinsic curvature, $K^{+a}\,_{a}$ and $K^{-a}\,_{a}$, it is found that
\begin{equation} 
-\frac{\epsilon}{8\pi}\left[	-\frac{\dot{\beta}_{+}}{\dot{R}}-\frac{\beta_{+}}{R} +\frac{\beta_{-}}{R}				\right]=0.
\label{A170}
\end{equation}
This is rewritten as
\begin{equation}
\frac{\dot{\beta}_{+}}{\dot{R}}=\frac{\beta_{-}}{R}-\frac{\beta_{+}}{R}. 
\end{equation}
Where, we have that $\beta_{+}=\dot{T}F$, $\dot{\beta}_{+}=F'\dot{R}\dot{T}$ and $\beta_{-}^{2}=1+\dot{R}^{2}$. Furthermore, we impose following condition
\begin{equation}
K^{-\tau}\,_{\tau}=0\longrightarrow \dot{\beta}_{-}=0.
\end{equation} 
So
\begin{equation}
\dot{\beta}_{+}=\frac{d}{d\tau}\left[\beta_{+}-\beta_{-}\right].
\end{equation}
So that, the equation (\ref{A170}) is simplificated to 
\begin{equation}
\frac{d}{d\tau}\left[\beta_{+}-\beta_{-}\right]=\frac{1}{R}\frac{dR}{d\tau}\left[\beta_{-}-\beta_{+} 		\right]
\label{A180}
\end{equation}
The integration of (\ref{A180}) is
\begin{equation}
\int \frac{d\left(\beta_{+}-\beta_{-}\right)}{\beta_{-}-\beta_{+}}=\int\frac{dR}{R}\longrightarrow C=R\left(\beta_{-}-\beta_{+}\right).
\label{A190}
\end{equation}
Well, if we are agree with (\ref{A160})
\begin{equation}
4\pi\sigma R^{2}=R\left(\beta_{-}-\beta_{+}\right)=C.
\end{equation}
In this point, we can define the rest mass of the Shell, $\mu=C$
\begin{equation}
\mu=R\left(\beta_{-}-\beta_{+}\right)\longrightarrow \beta_{+}=\beta_{-}-\frac{\mu}{R}.
\end{equation}
Getting the square root out of this and taking into account that $\beta_{-}^{2}=\dot{R}^{2}+1$, $\beta^{2}_{+}=\dot{R}^{2}+F$ and $F=1-\frac{2M}{R}$, where $M$ is the  gravitational mass and $M\neq \mu$. $M$ contains kinetic energy and potential gravitational for the Shell in gravitational contraction.

Therefore, we have
\begin{equation}
\beta_{+}^{2}=\beta_{-}^{2}-\frac{2\beta_{-}\mu}{R}+\frac{\mu^{2}}{R^{2}}=1+\dot{R}^{2}-\frac{2\mu}{R}\sqrt{1+\dot{R}^{2}}+\frac{\mu^{2}}{R^{2}}
\end{equation}
\begin{equation}
\frac{2M}{R}=\frac{2\mu}{R}\sqrt{1+\dot{R}^{2}}-\frac{\mu^{2}}{R^{2}},
\label{A200}
\end{equation}
the last equation is rewritten as
\begin{equation}
\frac{dR}{d\tau}=\sqrt{\left[\frac{M}{\mu}+\frac{\mu}{2R}\right]^{2}-1},
\end{equation}
we define the ratio between the gravitational mass $M$ and rest mass $\mu$ of the Shell as
\begin{equation}
a=\frac{M}{\mu},\,\,\,\, \frac{dR}{d\tau}=\sqrt{\left[a+\frac{M}{2aR}\right]^{2}-1}.
\label{AW210}
\end{equation} 

\section{The Shell kinematic}
In this appendix, we were able to draw the 3-position, 3-velocity and 3-acceleration of Shell, that is seen by an external observer in a flat and far asymptotic region for the Arenas-Castro solution
\begin{figure}[h]
\centering
	\includegraphics[width=0.7\textwidth]{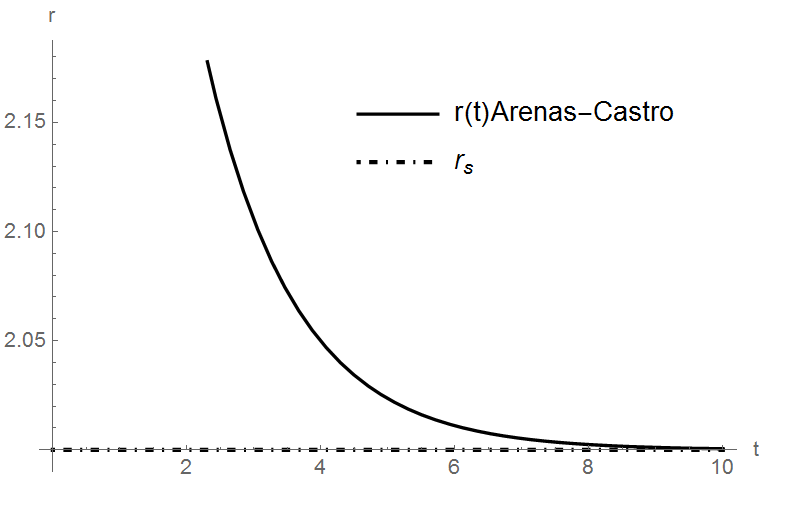}
\caption{The Shell 3-position to the equation (\ref{AB390}).}
\label{FA60}
\end{figure}
%
\begin{figure}[h]
\centering
	\includegraphics[width=0.7\textwidth]{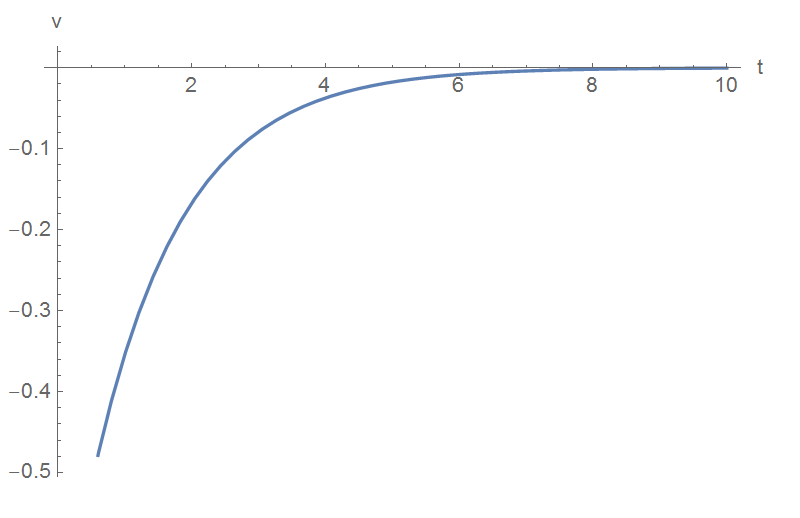}
\caption{The Shell 3-velocity.}
\label{FA70}
\end{figure}
%
\begin{figure}[h]
\centering
	\includegraphics[width=0.7\textwidth]{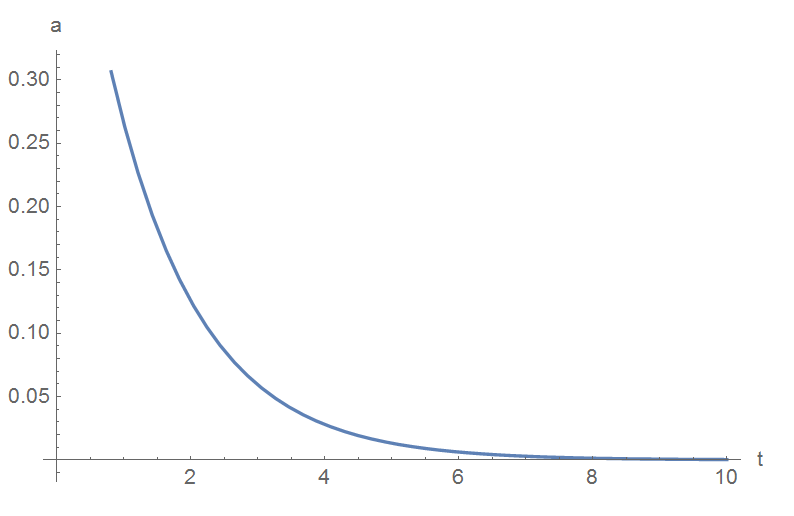}
\caption{The Shell 3-acceleration.}
\label{FA80}
\end{figure}
%


\section{Acknowledgments}
I want to thank Javier Cano and Walter Pulido for their valuable observations during the development of my research work.

This work was supported by  the Departamento de Administrativo de Ciencia, Tecnolog\'ia e Innovaci\'on, Colciencias.

\end{document}